# Computational ghost imaging with hybrid transforms by integrating Hadamard, discrete cosine, and Haar matrices

Yi-Ning Zhao[1], Lin-Shan Chen[1], Liu-Ya Chen[1], Lingxin Kong[1], Chong Wang[1,*], Cheng Ren[1], Su-Heng Zhang[2], and De-Zhong Cao[1,**]

[1]*Department of Physics, Yantai University, Yantai 264005, China*
[2]*Hebei Key Laboratory of Optic-Electronic Information and Materials, College of Physics Science & Technology, Hebei University, Baoding 071002, China*
*Email Address: wangchong@ytu.edu.cn
**Email Address: dzcao@ytu.edu.cn

**Abstract**: A scenario of ghost imaging with hybrid transform approach is proposed by integrating Hadamard, discrete cosine, and Haar matrices. The measurement matrix is formed by the Kronecker product of the two different transform matrices. The image information can be conveniently reconstructed by the corresponding inverse matrices. In experiment, six hybridization sets are performed in computational ghost imaging. For an object of staggered stripes, only one bucket signal survives in the Hadamard-cosine, Haar-Hadamard, and Haar-cosine hybridization sets, demonstrating flexible image compression. For a handmade windmill object, the quality factors of the reconstructed images vary with the hybridization sets. Sub-Nyquist sampling can be applied to either or both of the different transform matrices in each hybridization set in experiment. The hybridization method can be extended to apply more transforms at once. Ghost imaging with hybrid transforms may find flexible applications in image processing, such as image compression and image encryption.

## 1. Introduction

Hybrid transforms integrate the advantages of two or more different transforms. Rao et al. proposed the hybrid Hadamard-Haar transform for efficiently fast algorithms in image processing [1]. Petrosian investigated some classes of hybrid Hadamard-wavelet transforms for signal-image processing [2]. Sarukhanyan and Petrosian analyzed methods for construction and implementation of various parametric and hybrid wavelet transforms [3]. Recently, hybrid transforms have been widely used in image encryption, watermarking, and compression. Kumar et al. used the Kronecker product of two random matrices along with the double random phase encoding scheme to enhance the security in an optical image encryption system [4]. Selvam et al. proposed a hybrid transform-based reversible watermarking technique in telemedicine applications without any additional key information [5]. Thakur et al. used a hybrid transform of wavelet packet transform and block-discrete-cosine transform to achieve high-quality image compression [6]. Thayammal et al. proposed a hybrid transform-based multispectral compression to preserve image edges [7]. Sheela et al. exploited a hybrid random matrix transform to perform confusion operation which is controlled by 5D hyperchaotic system, and performed image encryption [8].

Computational ghost imaging (CGI) [9] uses a bucket detector to collect the object beam. Image reconstruction is performed by correlating the (known) random projected patterns and the bucket detection signals. In conventional ghost imaging experiments, the measurement matrix is composed of random patterns [9-12]. The orthogonal transform matrices are used in recent works in CGI. Zhang et al. [13,14] utilized Fourier transform bases to reconstruct high-quality images. Walsh-Hadamard bases, in normal order, cake-cutting order, Russian doll order, and origami order, as a measurement matrix were widely investigated [15-18]. Haar wavelet base [19] and other wavelet [20] were applied to accurately extract the target information. These orthogonal transforms have greatly improved the image quality and reconstruction speed. However, ghost imaging with a single transform is limited in complex image processing. In this paper, we propose an approach of ghost imaging with hybrid transforms (GIHT) to exploit the transform hybridization.

We adopt Hadamard, discrete cosine, and Haar transform matrices in GIHT. The measurement matrix is formed by the Kronecker products of the hybrid matrices. The images are conveniently reconstructed with the corresponding inverse transform matrices. For an object of staggered stripes, only one bucket signal is nonzero in Hadamard-cosine, Haar-Hadamard, and Haar-cosine hybridization sets, demonstrating flexible image compression. For a handmade windmill object, the peak signal-to-noise (PSNR) and structural similarity (SSIM) of the reconstructed image vary with the hybridization sets. Sub-Nyquist sampling is applied to either or both of the two hybrid transform matrices. A non-square measurement matrix is formed by the Kronecker product of the truncated transform matrices. Generally, the hybridization method can be extended to the case of three and more transform matrices at once. The measurement matrix can be obtained by sequential Kronecker products of these transform matrices.

## 2. Theory

In CGI, a sequence of random patterns from a thermal light are projected onto the object. The object beam is collected by a bucket detector, then converted into a series of bucket signals

$$\mathbf{y}_{M\times 1} = \mathbf{A}_{M\times N}\mathbf{x}_{N\times 1}, \tag{1}$$

where the vector $\mathbf{x}$ is the one-dimensional distribution of the object, and the measurement matrix $\mathbf{A}$ is composed of $M$ row vectors. For simplicity, the footnotes, which indicate the matrix sizes, are ignored below. The image can be reconstructed by the intensity correlation function between the bucket signals and random patterns that

$$\mathbf{x}' = \mathbf{A}^\dagger \mathbf{y}, \tag{2}$$

where $\dagger$ represents matrix transpose conjugation.

The reconstructed image by Eq. (2) suffers from low contrast, due to the inevitable constant background [21]. Many efforts were made to improve the reconstructed images in early investigations of ghost imaging with thermal light, such as differential method [22], high-order correlation function [23], and compressed sensing [10].

*2.1 Review of Fourier CGI*

Recently, orthogonal transforms [13-17] have been widely adopted in ghost imaging. With Fourier transforms [13,14], the ghost imaging system reduces time, and improves the image quality. The matrix elements of the discrete Fourier transform are

$$F_{n,n'} = \frac{1}{\sqrt{N}} e^{i\frac{2\pi nn'}{N}}, \tag{3}$$

where $n, n' = 0, 1, 2, \cdots, N-1$. The bucket signals in Eq. (1) and image reconstruction in Eq. (2) are now written as $\mathbf{y} = \mathbf{F}\mathbf{x}$, and $\mathbf{x} = \mathbf{F}^\dagger \mathbf{y}$, respectively.

In general, we assume that the object/image is distributed over a range of $M\times N$ pixels. By replacing the Fourier transform matrix $\mathbf{F}$ with two smaller matrices $\mathbf{F}_1$ and $\mathbf{F}_2$, the bucket signals and reconstructed image in matrix form are written as

$$\mathbf{Y} = \mathbf{F}_1 \mathbf{X} \mathbf{F}_2^\dagger, \quad \mathbf{X} = \mathbf{F}_1^\dagger \mathbf{Y} \mathbf{F}_2, \tag{4}$$

where $\mathbf{Y}$ and $\mathbf{X}$ are two-dimensional. The extended theory is also valid for Hadamard CGI [15-17, 19].

Let us take an object/image of resolution $32\times 64$ as an example to estimate the RAM requirements in image reconstruction. In one-dimensional case of Eq. (1), the vector of the object/image $\mathbf{x}$ is composed of $32\times 64 = 2,048$ pixels. The measurement matrix $\mathbf{F}$ and reconstruction matrix $\mathbf{F}^\dagger$ are both composed of $2048\times 2048 = 4,194,304$ pixels. In two-dimensional case of Eq. (4), however, $\mathbf{F}_1$ is of size $64\times 64 = 4,096$, and $\mathbf{F}_2$ is of size $32\times 32 = 1,024$. In comparison, the RAM requirement and time consumed for image reconstruction in two-dimensional case can be greatly reduced.

*2.2 GIHT theory*

In this paper, we propose a scenario of ghost imaging with hybrid orthogonal transforms, by generalizing the mechanism of Fourier ghost imaging. By replacing $\mathbf{F}_1$ and $\mathbf{F}_2$ with different orthogonal transform matrices $\mathbf{L}$ and $\mathbf{R}$, we rewrite Eq. (4) in a general form as

$$\mathbf{L}\mathbf{X}\mathbf{R}^\dagger = \mathbf{Y}. \tag{5}$$

For two-dimensional images, $\mathbf{L}$ and $\mathbf{R}$ can be different transforms of different sizes.

Due to the matrix orthogonality, the reconstructed image is obtained from

$$\mathbf{X} = \mathbf{L}^\dagger \mathbf{Y} \mathbf{R}, \tag{6}$$

which will produce a perfect image if there is no ambient noise in the imaging system.

With respect to Eq. (5), we have two statements.

**Statement I:** The measurement matrix is $\mathbf{A} = \mathbf{L} \otimes \mathbf{R}$.

Proof: According to the properties of Kronecker products, Eq. (5) can be rewritten as

$$vec(\mathbf{L}\mathbf{X}\mathbf{R}^\dagger) = (\mathbf{L} \otimes \mathbf{R})vec(\mathbf{X}), \tag{7}$$

where the vectorization

$$vec(\mathbf{X}) = [X_{11}, X_{12}, \cdots, X_{1N}, X_{21}, X_{22}, \cdots, X_{MN}]^T \tag{8}$$

is to stack the rows of matrix $\mathbf{X}$ to form a column vector, $T$ stands for matrix transpose. Combining Eqs. (5) and (7), we obtain the standard form of ghost imaging

$$(\mathbf{L} \otimes \mathbf{R})\mathbf{x} = \mathbf{A}\mathbf{x} = \mathbf{y}, \tag{9}$$

where $\mathbf{x} = vec(\mathbf{X})$, $\mathbf{y} = vec(\mathbf{Y})$. Equation (9) is the same one-dimensional case as Eq. (1).

**Statement II:** Matrix $\mathbf{A}$ is orthogonal since $\mathbf{L}$ and $\mathbf{R}$ are orthogonal.

Proof: For the orthogonal matrices $\mathbf{L}$ and $\mathbf{R}$, we obtain the inner products of their row vectors

$$\sum_n R_{n'n}R_{n''n} = \delta_{n'n''}, \quad \sum_m L_{m'm}L_{m''m} = \delta_{m'm''}, \tag{10}$$

where the Dirac delta function is $\delta_{n'n''} = 1$ for $n' = n''$ and $\delta_{n'n''} = 0$ for $n' \neq n''$. Then we obtain the inner product of two different vectors of the measurement matrix

$$\sum_l A_{l'l}A_{l''l} = \delta_{l'l''}, \tag{11}$$

which insures matrix orthogonality of $\mathbf{A}$.

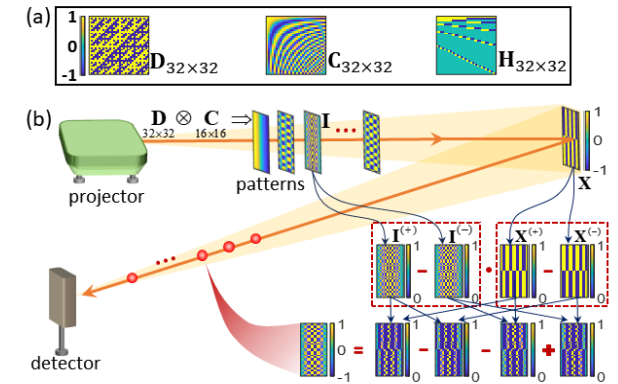

Fig. 1. (a) Hadamard (D), DCT (C), and Haar (H) matrices. (b) Experimental setup.

The orthogonality of the measurement matrix will be beneficial for image reconstruction and compression. In experiment, each row of $\mathbf{A}$ is reshaped to a projected pattern which is of the same size as $\mathbf{X}$, and then is projected onto the object. The $k$-th projected pattern can be obtained by

$$\mathbf{I}_k = vec[\mathbf{L}(m,:)]\{vec[\mathbf{R}^\dagger(:,n)]\}^T, \tag{12}$$

where $k = (m-1)\times N + n$.

## 2.3 Three matrices

Three orthogonal matrices, Hadamard, discrete cosine transform (DCT) and Haar matrices, are adopted in GIHT. The three matrices of $32\times 32$ are shown in Fig. 1(a).

Specifically, the 2-order and $2^n$-order Hadamard matrices are expressed as

$$\mathbf{D}_{2\times 2} = \frac{1}{\sqrt{2}}\begin{pmatrix} 1 & 1 \\ 1 & -1 \end{pmatrix}, \quad \mathbf{D}_{2^n\times 2^n} = \frac{1}{\sqrt{2}}\begin{pmatrix} \mathbf{D}_{2^{n-1}} & \mathbf{D}_{2^{n-1}} \\ \mathbf{D}_{2^{n-1}} & -\mathbf{D}_{2^{n-1}} \end{pmatrix}, \quad (13)$$

where $n = 2, 3, \cdots$.

The DCT matrix $\mathbf{C}$ consists of elements as

$$C_{m'm''} = c_{m'}\cos\left(m'\pi\frac{m''+0.5}{M}\right), \quad (14)$$

where the coefficients $c_0 = 1/\sqrt{M}$, $c_{m>0} = \sqrt{2/M}$, and $m', m'' = 0, 1, 2, \cdots, M-1$.

The Haar matrix is composed of Haar bases as

$$\mathbf{H}^T_{2^n\times 2^n} = \left(w_1, h_0^0, h_0^1, h_1^1, h_0^2, h_1^2, h_2^2, h_3^2, \cdots, h_k^j, \cdots, h_{2^{n-1}-1}^{n-1}\right), \quad (15)$$

where the $2^n$-dimensional (column) vectors are

$$w_1 = \begin{pmatrix} 1 & 1 & \cdots & 1 \end{pmatrix}^T, \quad (16)$$

$$h_k^j(i) = \begin{cases} 0, & 1 \le i \le k\times 2^{n-j}, \\ 1, & k\times 2^{n-j}+1 \le i \le k\times 2^{n-j}+2^{n-j-1}, \\ -1, & k\times 2^{n-j}+2^{n-j-1}+1 \le i \le (k+1)\times 2^{n-j}, \\ 0, & (k+1)\times 2^{n-j}+1 \le i \le 2^n, \end{cases} \quad (17)$$

with $0 \le j \le n-1$, $0 \le k \le 2^j-1$, and $i \in (1, 2, \cdots, 2^n)$.

## 3. Experimental results of image compression

The setup of GIHT experiment is depicted in Fig. 1(b). A projector (LG: PF1500G-GL) is used to play patterns out of the measurement matrix. The reflected object beam is collected by a charge coupled device (Mintron: MTV-1881EX), which serves as a bucket detector. The projected patterns and bucket signals are processed in a computer for image reconstruction.

In practice, the values of the projected patterns are normalized to [-1, 1]. Each pattern $\mathbf{I}$ is represented by two positive patterns $\mathbf{I}^{(+)} = (1+\mathbf{I})/2$ and $\mathbf{I}^{(-)} = (1-\mathbf{I})/2$ that $\mathbf{I} = \mathbf{I}^{(+)} - \mathbf{I}^{(-)}$.

A binary (-1 and 1) virtual object of staggered stripes, as shown in Fig. 1(b) and Fig. 2(a), is loaded to the projector along with the projected patterns. Similarly, the object function is represented as $\mathbf{X} = \mathbf{X}^{(+)} - \mathbf{X}^{(-)}$, where the two positive patterns are $\mathbf{X}^{(+)} = (1+\mathbf{X})/2$ and $\mathbf{X}^{(-)} = (1-\mathbf{X})/2$. That is, it takes four projections to get a bucket detection signal, which is the sum of $\mathbf{I}\cdot\mathbf{X} = [\mathbf{I}^{(+)} - \mathbf{I}^{(-)}]\cdot[\mathbf{X}^{(+)} - \mathbf{X}^{(-)}]$, where the dot means Hadamard product. The subplots in Fig. 1(b) show the processes of pattern representation and projection, where the object of $32\times 16$ pixels and the measurement matrix $\mathbf{A} = \mathbf{D}_{32\times 32}\otimes\mathbf{C}_{16\times 16}$ are exhibited.

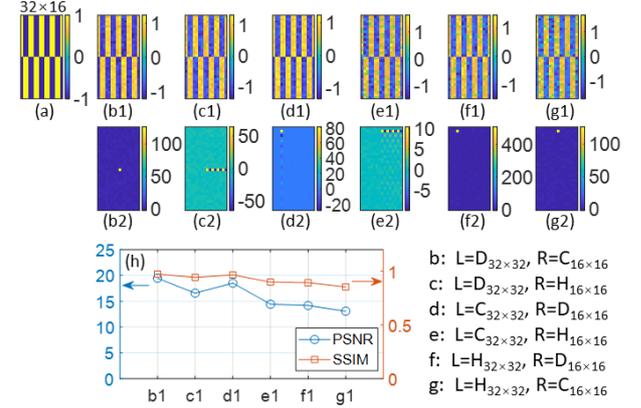

Fig. 2. Experimental results with an object of staggered stripes. (a) The virtual object. (b1-g1) The reconstructed images. (b2-g2) The corresponding bucket detection signals. (h) The PSNRs and SSIMs of the images. The hybrid matrices are specified.

The specific object $\mathbf{X}$ ($32\times 16$) is shown in Fig. 2(a). The images in Figs. 2(b1), 2(c1), 2(d1), 2(e1), 2(f1), and 2(g1) are reconstructed with hybridization sets of $\mathbf{L} = \mathbf{D}_{32\times 32}$, $\mathbf{R} = \mathbf{C}_{16\times 16}$; $\mathbf{L} = \mathbf{D}_{32\times 32}$, $\mathbf{R} = \mathbf{H}_{16\times 16}$; $\mathbf{L} = \mathbf{C}_{32\times 32}$, $\mathbf{R} = \mathbf{D}_{16\times 16}$; $\mathbf{L} = \mathbf{C}_{32\times 32}$, $\mathbf{R} = \mathbf{H}_{16\times 16}$; $\mathbf{L} = \mathbf{H}_{32\times 32}$, $\mathbf{R} = \mathbf{D}_{16\times 16}$; and $\mathbf{L} = \mathbf{H}_{32\times 32}$, $\mathbf{R} = \mathbf{C}_{16\times 16}$, respectively. The corresponding bucket detection signals $\mathbf{Y}$ are shown in Figs. 2(b2), 2(c2), 2(d2), 2(e2), 2(f2), and 2(g2). The PSNRs and SSIMs [24] of these reconstructed images are plotted by the blue circles and orange squares in Fig. 2(h). We can see that all the reconstructed images are of good quality. The PSNR and SSIM of the image in Fig. 2(b1) reach their maximums $PSNR = 19.45$ and $SSIM = 0.97$, and those in Fig. 2(g1) arrive at their minimums $PSNR = 13.07$ and $SSIM = 0.85$. The image quality fluctuates slightly due to the uncontrollable ambient noise.

In Figs. 2(b2), 2(f2) and 2(g2), an interesting feature is that only one peak exists in each pattern of the bucket signals $\mathbf{Y}$, and other bucket detection signals are almost zero. Indeed, the theoretical values are exact zeros. This feature indicates that GIHT is flexible for image compression.

## 4. GIHT experiment with a real object

In what follows, we perform the GIHT experiment of image reconstruction with a real object, a handmade windmill object as shown in Fig. 3(a). The image resolution is $32\times 64$.

### 4.1 GIHT under full sampling

Similar to that in Fig. 2, the hybridization sets for the images in Figs. 3(b), 3(c), 3(d), 3(e), 3(f), and 3(g) are

$L = D_{32 \times 32}$, $R = C_{64 \times 64}$; $L = D_{32 \times 32}$, $R = H_{64 \times 64}$; $L = C_{32 \times 32}$, $R = D_{64 \times 64}$; $L = C_{32 \times 32}$, $R = H_{64 \times 64}$; $L = H_{32 \times 32}$, $R = D_{64 \times 64}$; and $L = H_{32 \times 32}$, $R = C_{64 \times 64}$, respectively. The PSNRs and SSIMs of all the areas in the red frames of the images are plotted by the blue line with circles and the orange line with squares in Fig. 3(h).

All the SSIM values are very small since the corresponding area of the object in the red frame is empty. The maximum ($6.346 \times 10^{-5}$) and minimum ($0.32 \times 10^{-5}$) of SSIM values exist in Figs. 3(d) and 3(g), respectively. And the maximum ($13.58$) and minimum ($6.96$) of PSNR values exist in Figs. 3(b) and 3(e), respectively.

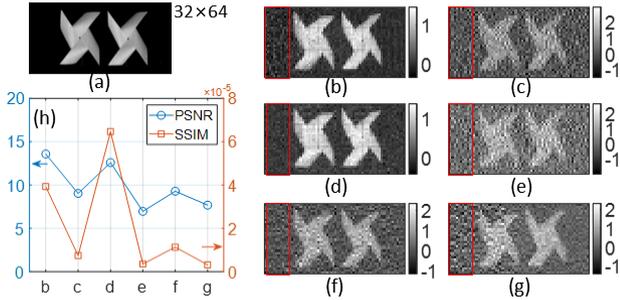

Fig. 3. Experimental results in the full-sampling case. (a) Object of handmade windmill. (b-g) The reconstructed images with different hybrid matrices. (h) The PSNR and SSIM of the areas inside the red frames in all the reconstructed images.

*4.2 Sub-Nyquist sampling*

In the above experimental results, full sampling is performed that $L_{M \times M} X_{M \times N} R^{\dagger}_{N \times N} = Y_{M \times N}$. The measurement matrix is of size $MN \times MN$ that $A_{MN \times MN} = L_{M \times M} \otimes R_{N \times N}$. However, this does not mean full sampling is the only choice.

The sub-Nyquist sampling can be applied in GIHT by under-sampling the two transform matrices that $L_{M \times M} \to L_{M_L \times M}$, $R_{N \times N} \to R_{N_R \times N}$, where $M_L < M$ and $N_R < N$. Therefore, under the condition of sub-Nyquist sampling, the measurement matrix becomes
$$A_{M_L N_R \times MN} = L_{M_L \times M} \otimes R_{N_R \times N}, \tag{18}$$
and the bucket signals are
$$Y_{M_L \times N_R} = L_{M_L \times M} X_{M \times N} R^{\dagger}_{N \times N_R}. \tag{19}$$
The image can then be reconstructed by
$$X_{M \times N} = L^{\dagger}_{M \times M_L} Y_{M_L \times N_R} R_{N_R \times N}. \tag{20}$$

In experiment, the sampling parameters are chosen as $M_L = 0.906M = 29$ and $N_R = 0.906N = 58$ for the Hadamard, cosine, and Haar matrices in the hybridization sets in experiment. The total sampling rate is about $M_L N_R / MN = 0.821$. The reconstructed images in the sub-Nyquist sampling case are shown in Fig. 4. The hybridization sets for the images in Figs. 4(a)-4(f) are the same as that in Figs. 3(b)-3(g). The PSNRs and SSIMs of all the areas in the red frames of the images are plotted by the blue line with circles and the orange line with squares in Fig. 4(g).

In the sub-Nyquist sampling case, the quality factors of the reconstructed images in Fig. 4 are quite well, compared with that in Fig. 3. The PSNR maximum ($13.66$) and minimum ($7.70$) of the reconstructed images are in Figs. 4(a) and 4(f), respectively. And the SSIM maximum ($5.23 \times 10^{-5}$) and minimum ($0.41 \times 10^{-5}$) of the reconstructed images are in Figs. 4(c) and 4(d), respectively.

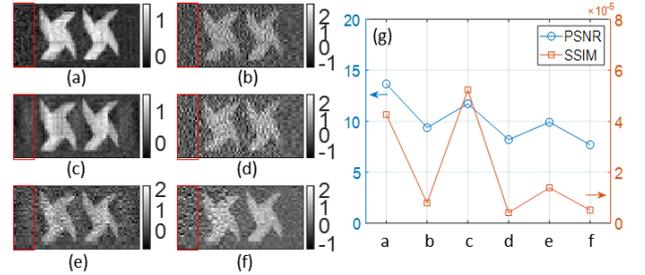

Fig. 4. Experimental results in the sub-Nyquist sampling case. (a-f) The reconstructed images with different hybrid matrices. (g) The PSNR and SSIM of the areas inside the red frames in all the reconstructed images.

There is not much difference between the quality factors of the images in the full sampling and sub-Nyquist sampling cases in Figs. 3 and 4, due to the large sampling rate. The image quality will become worse if the sampling rate is low.

## 5. Discussion and conclusion

Anyway, ghost imaging with two hybrid transform matrices are illustrated. By extending the idea, GIHT can be implemented with many more transform matrices. Obviously, the extended version of GIHT in Eq. (5) can be directly written as
$$L^{(j)} \cdots L^{(2)} L^{(1)} X [R^{(1)}]^{\dagger} [R^{(2)}]^{\dagger} \cdots [R^{(k)}]^{\dagger} = Y, \tag{21}$$
where $j$ matrices on the left side and $k$ matrices on the right side of $X$. The measurement matrix in experiment can be obtained by
$$A = [L^{(1)} \otimes R^{(1)}][L^{(2)} \otimes R^{(2)}] \cdots [L^{(k)} \otimes R^{(k)}], \tag{22}$$
where $j = k$ has been considered. In the case when $j \neq k$, identity matrices should be supplied. The image reconstruction can be implemented by
$$[L^{(1)}]^{\dagger} [L^{(2)}]^{\dagger} \cdots [L^{(j)}]^{\dagger} Y R^{(k)} \cdots R^{(2)} R^{(1)} = X. \tag{23}$$
In fact, GIHT can be performed with any hybrid transform matrices, no matter unitary, orthogonal, or random. Since many transforms can be applied, the transform types and matrix positions provide convenience for image processing, such as image compression, encryption, encoding, etc.

In summary, we have proposed a scenario of ghost imaging with hybrid transform matrices in two-dimensional case. In experiment, three orthogonal transform matrices (Hadamard, DCT, and Haar) have been adopted. The dimensions of the two matrices do not need to be the same, they can just match the size of the 2D images. The measurement matrix is obtained by the Kronecker products of the two hybrid transform matrices. GIHT can exploit the characteristics of the hybridization. For an object of staggered stripes, convenient image compression is obtained since only one bucket detection signal survives in several hybridization sets. Image reconstruction and sub-Nyquist sampling are investigated for a handmade windmill object. The image PSNR and SSIM vary with the hybridization sets, due to the ambient noise. Therefore, GIHT can find its wide applications in image processing.

**Declaration of Competing Interest**

The authors declare that they have no known competing financial interests or personal relationships that could have appeared to influence the work reported in this paper.

**CRediT authorship contribution statement**

Yi-Ning Zhao: Performing experiment, **Lin-Shan Chen**: Software, **Liu-Ya Chen**: Performing experiment, Writing- original draft, **Lingxin Kong**: Writing –review & editing, **Chong Wang**: Writing –review & editing, Software, **Cheng Ren**: Writing –review & editing, **Su-Heng Zhang**: Writing –review & editing, Software, and **De-Zhong Cao**: Writing –original draft, review & editing.

**Acknowledgements**

This work is supported by the National Natural Science Foundation of China (No. 62105278 & 11674273), and the Natural Science Foundation of Hebei province (No. A2022201039).